# Passive Vibration Isolation Characteristics of Negative Extensibility Metamaterials

Somya Ranjan Patro[a], Hemant Sharma[a], Seokgyu Yang[a] and Jinkyu Yang[a,*]

[a] Department of Mechanical Engineering, Seoul National University, Seoul, South Korea

| ARTICLE INFO | ABSTRACT |
|---|---|
| *Keywords*:<br><br>*Negative extensibility, Passive vibration isolation, Braess paradox, Countersnapping instability, Mechanical metamaterial* | Negative extensibility refers to the category of mechanical metamaterials having an unusual phenomenon where the system contracts upon expansion. The dynamic analysis of such systems is crucial for exploring the vibration isolation characteristics, forming the prime focus of the present study. Inspired by Braess's paradox, the mechanical model incorporates coupled tunable nonlinear spring stiffness properties (strain hardening and softening), which alternate when a certain displacement threshold is exceeded. This stiffness-switching mechanism facilitates low-frequency passive vibration isolation using the phenomenon of countersnapping instability. The vibration isolation characteristics resulting from the stiffness-switching mechanism are investigated using time- and frequency-domain plots. Furthermore, the relationship between the stiffness switching mechanism and various system parameters is visualized using a three-dimensional parametric space. The efficacy of the proposed system is evaluated by comparing it with the existing bi-stable systems, revealing superior performance in isolating high-amplitude vibrations. The proposed mechanism enhances the understanding of dynamic behaviors in critical structural elements for multi-stable mechanical metamaterials, providing insights and opportunities for innovative adaptive designs. |

## 1. Introduction

Mechanical metamaterials are the architectured materials whose properties are derived from engineered internal structures rather than chemical composition of the bulk material, exhibiting novel characteristics uncommon in most natural and conventional artificial structures [1–3]. These unconventional properties contradict intuitive expectations, including negative Poisson's ratio, negative stiffness, negative compressibility/extensibility, etc. [4,5]. Leveraging these distinctive attributes, mechanical metamaterials have attracted significant interest in diverse fields such as energy harvesting, light-weight deployable structures, vibration isolation, fault detection, and soft robots, to name a few. [6,7].

Among the numerous exotic properties of mechanical metamaterials [8], recently, intriguing properties like negative extensibility/compressibility have increasingly drawn attention from the research community due to their potential in adaptive, self-tuning vibration isolation and smart material applications [9]. Negative compressibility/extensibility refers to the peculiar phenomenon where materials expand upon compression or contract upon decompression [10]. Initially, such a system deforms in a typical manner: expanding under tension and contracting under compression. However, once the applied load surpasses a critical threshold, this behavior reverses: the system exhibits significant deformation in the direction opposite to the applied load [11].

A classic example of this phenomenon is Braess's paradox, or the *spring paradox* [12], where a network composed of springs and strings paradoxically elevates a weight at equilibrium when a supporting string is severed [13] (see Fig. 1(a)). This counterintuitive effect arises due to alterations in load paths and mechanical impedance mismatches [14]. Inspired by this concept, various studies have explored bistable unit cells, stress-induced metastable state transitions, and capillary-driven phase transformations to achieve unconventional mechanical responses [15–17]. Furthermore, experimental validations, phase diagrams, and advanced design strategies, including model replacement methods, have been developed to optimize these metamaterial effects [9,11]. Despite these advancements, there is a persistent lack of thorough research on the dynamic characteristics of negative extensibility/compressibility systems.

Negative extensibility (NE) metamaterials can exhibit intriguing dynamic behaviors due to their tunable stiffness, structural nonlinearity, and phase transition mechanisms. These properties make them promising candidates for applications such as adaptive vibration isolation for a wide-frequency range vibration. In line with this, conventional nonlinear vibration isolators, such as bistable [18] and quasi-zero stiffness (QZS) systems [19], have been thoroughly investigated for their ability to manage low-frequency vibrations effectively. However, QZS systems are typically monostable with a single equilibrium point and may suffer from stiffness bifurcation or

---

dynamic instability under large-amplitude excitation, which can compromise their reliability and performance [20]. Similarly, bistable systems can undergo inter-well oscillations when subjected to high excitation levels, potentially introducing large forces or displacements that may damage the primary structure [21].

In contrast to conventional nonlinear systems, negative extensibility systems present a unique advantage, including controllable mode switching and stiffness modulation under dynamic loading. These features can, in return, alter the system's natural frequencies without requiring any external input. This motivates us to perform a comprehensive theoretical investigation into the dynamic behavior of NE metamaterials, aiming to evaluate their effectiveness and explore their potential applications in efficient vibration isolation.

A recent study demonstrated that externally driven mass-spring systems can achieve resonance avoidance through countersnapping instability (CSI), enabling passive self-switching of the system's natural frequencies, a mechanism inspired by the classical spring paradox [9]. The core principle behind NE systems lies in their distinctive mode-switching behavior during load–unload cycles, which opens new possibilities for harnessing these dynamic characteristics. Despite the promise in these initial demonstrations, a thorough investigation into the dynamics of the NE metamaterials remains lacking and needs further exploration.

Motivated by these insights, the present study aims to investigate the vibration isolation capabilities of NE systems through a comprehensive generalized analytical approach. The primary novelty lies in systematically modeling a mechanical metamaterial exhibiting NE characteristics and thoroughly analyzing its vibration isolation performance under dynamic loading conditions. The system consists of a tunable nonlinear spring stiffness capable of switching from strain-hardening (SH) to strain-softening (SS) behavior via CSI mechanisms upon exceeding a specific threshold. The effectiveness of passive vibration isolation has been assessed through detailed frequency-domain and time-domain analyses. Additionally, a comprehensive three-dimensional parametric space examines the sensitivity of CSI behavior to dynamic properties, including damping, amplitude, and excitation frequency. Finally, the proposed NE system's effectiveness is systematically compared with generalized bistable systems, quantifying the vibration isolation performance across various excitation amplitudes. It is observed that the proposed system demonstrates superior vibration isolation performance in both the transient and steady-state phases, especially at high amplitude vibration, compared to the bi-stable systems.

## 2. The concept of Spring (Braess) Paradox

The present study is inspired by the spring (Braess) paradox mechanism [14] which uses a spring-string model shown in Fig. 1(a). The system comprises an object of weight $W$ that is supported by two identical springs (stiffness $k$ each) and three strings (one primary and two secondary). Initially, the secondary strings are slack and only the primary string and the springs bear the load, forming a series spring connection with effective stiffness $k_{eff} = k/2$. In this arrangement, the object experiences an initial displacement $\Delta u_i$ due to self-weight (Fig. 1(a), left), the restoring force of the system can be determined as $F = k/2 \times \Delta u_i$. Now, upon severing the primary string, the secondary strings carry the load, transforming the configuration into a parallel arrangement (Fig. 1(a), right), which increases the system stiffness to $k_{eff} = 2k$, four times that of the previous series configuration. Under the same applied load, the total displacement reduces to $\Delta u_f = \Delta u_i/4$, so that the net contraction of the object is $\Delta u = \Delta u_i - \Delta u_f = 3\Delta u_i/4$. This behavior is denoted as the negative extensibility (NE) phenomenon.

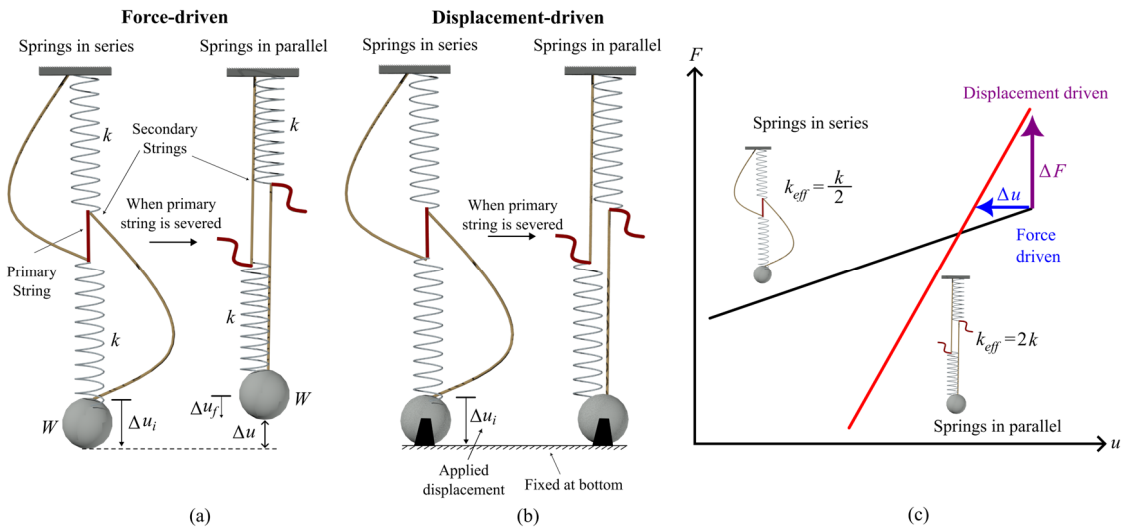

**Fig. 1.** The spring (Braess) paradox concept using the spring-string model [14]. An object of weight $W$ is connected by two linear springs of equal stiffness $k$ and three strings (one primary string and two secondary strings). The two secondary strings do not contribute to carrying the object, and the two springs, along with the primary string, are connected in series. (a) In a force-driven approach, severing the primary string will result in the descent rather than the ascent of the weight due to the contribution of the secondary strings in carrying the weight, thus demonstrating irreversible negative extensibility (NE). (b) In the displacement-driven approach, the bottom of the weight is fixed to the ground, and severing the primary string increases the system stiffness, resulting in a larger restoring force for the same displacement, thus demonstrating counter-snapping instability (CSI). (c) Force-displacement plot for the spring paradox system showing both force-driven (negative extensibility) and displacement-driven (increased restoring force) approaches.

In the displacement-controlled case (Fig. 1(b)), the object is first elongated by $\Delta u_i$ and held fixed. Severing the primary string again shifts the configuration to parallel, but with the ends fixed, the load jumps fourfold while the displacement remains fixed at $\Delta u_i$. This phenomenon is called counter-snapping instability (CSI). The behavior of both force-driven and displacement-driven approaches is shown in the force-displacement plot of Fig. 1(c), where the black solid line represents the series connection $(k_{eff} = k/2)$ and red solid line represents a parallel connection $(k_{eff} = 2k)$. Initially, the object's displacement follows the black solid line up to a certain threshold, then it jumps to the red solid line while transforming from a series to a parallel arrangement. For the force-driven case, this transformation yields a negative displacement of $\Delta u$ due to the free boundary condition and for the displacement-driven case, it yields a sudden force jump $(\Delta F)$ due to being fixed at both ends. For better visualization, an animation of the spring paradox mechanism considering both force and displacement-driven approaches, along with the respective force-displacement plot, is provided in the supplementary material (Appendix A. Video S1). In the later sections, the dynamics of such NE mechanisms are investigated, taking into account generalized nonlinear behavior representing both series and parallel configurations.

## 3. Mathematical formulation

The dynamic model of a standard spring paradox mechanism can be idealized as a single spring-mass-damper system having tunable spring stiffness behavior, as shown Fig. 2(a). To have a generalized dynamic analysis, this study considers nonlinear stiffness behavior to represent the spring elements in the spring paradox mechanism; specifically, strain hardening and strain softening stiffness behavior for series and parallel arrangements, respectively. The negative extensibility (NE) mechanism is designed so that a counter-snapping instability (CSI) phenomenon is observed when the dynamic amplitude response exceeds a certain threshold. Thus, the overall stiffness of the system switches from strain hardening (SH) to strain softening (SS) as shown in Fig. 2(b). For simplicity, the non-linear behavior of SH and SS is represented by cubic non-linearity. Thus, the differential equation of motion [22] for the system subjected to a harmonic force excitation, shown in Fig. 2(a), can be expressed as follows:

$$m\ddot{x} + c\dot{x} + \kappa_1 x + \kappa_3 x^3 = F_0 \cos(\omega t) \tag{1}$$

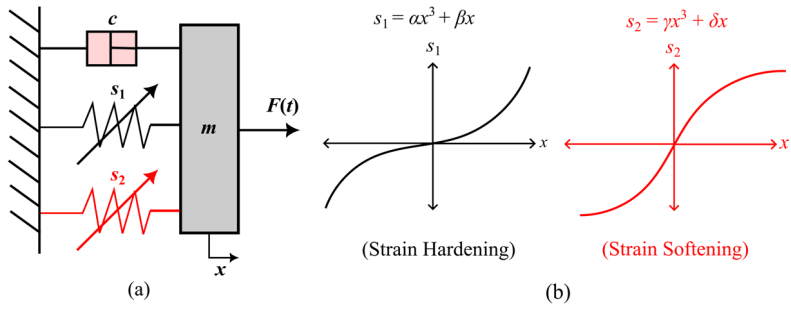

**Fig. 2.** (a) Illustration of the proposed spring-mass-damper system having tunable non-linear strain hardening (SH) to strain softening (SS) stiffness switching ability, subjected to a harmonic force excitation. (b) The corresponding force-displacement profile for SH and SS behavior is represented by cubic non-linearity.

where $m$ is the mass, $c$ represents the linear damping, $\kappa_1 x + \kappa_3 x^3$ is the cubic non-linear restoring force assumed for the system, which can be either SH ($\kappa_1 = \beta, \kappa_3 = \alpha$) or SS ($\kappa_1 = \delta, \kappa_3 = \gamma$) as shown in Fig. 2(b), and $F_0 \cos(\omega t)$ is the harmonic force excitation of amplitude $F_0$ and frequency $\omega$. Throughout the present study, the values of $\alpha, \beta, \delta$ and $\gamma$ are chosen, such as $\gamma = -\alpha$ and $\delta = 4\beta$ to represent the series and parallel connection of spring paradox, and to maintain uniformity. Eq. (1) can be nondimensionalized by introducing several parameters:

$$\tilde{x} = \frac{x}{X}, \lambda = \frac{\omega}{\omega_n}, \omega_n = \sqrt{\frac{\kappa_1}{m}}, \tau = \omega_n t, \xi = \frac{c}{2m\omega_n}, \tilde{\kappa} = \frac{X^2 \kappa_3}{\kappa_1}, \Gamma = F_0 \sqrt{\frac{\kappa_3}{\kappa_1^3}} \tag{2}$$

where $\omega_n$ is the natural frequency corresponding to the linear stiffness component, $\tilde{x}, \lambda, \xi, \tau$ are the dimensionless displacement, the frequency ratio, the damping ratio, and the dimensionless time, respectively. $\tilde{\kappa}$ and $\Gamma$ are the dimensionless non-linear stiffness coefficient and forcing amplitude. $X$ is the basic displacement for dimensionless processing, which is assumed as 1 in the present study. Substituting Eq. (2) in Eq. (1), the non-dimensional equation of motion can be expressed as follows:

$$\tilde{x}'' + 2\xi \tilde{x}' + \tilde{s} = \Gamma \cos(\lambda \tau) \tag{3}$$

where $\tilde{s} = \tilde{x} + \tilde{\kappa}\tilde{x}^3$. The approximate analytical solution of this nonlinear system is approximated by the harmonic balance method (HBM) [22–24]. The harmonic solution of the system can be expressed as follows:

$$\tilde{x}(\tau) = \tilde{X}\cos(\lambda\tau - \phi) \quad (4)$$

where $\tilde{X}$ is the dimensionless displacement amplitude and $\phi$ is the phase lag. Substituting Eq. (4) in Eq. (3) and neglecting higher order cosine terms such as, $\cos(3\lambda\tau)$, the frequency-amplitude relationship can be obtained (see supplementary materials Appendix C. for detailed derivations). The subsequent sections explore the stiffness switching behavior using the counter-snapping instability (CSI) mechanism.

## 4. Dual stiffness behavior via Counter Snapping Instability (CSI)

Since the natural frequencies of nonlinear systems vary with initial and excitation conditions, the present study excites the system based on the linear tangent stiffness of the initial configuration ($\kappa_1$). When the displacement amplitude of the system reaches a certain pre-defined threshold $\tilde{x}_T$, the non-linear stiffness of the system is switched from strain hardening (SH) to strain softening (SS) due to the Counter Snapping Instability (CSI) phenomenon, as shown in the force displacement curve of Fig. 3(a). The stiffness switching can be either force-driven (displacement gets reduced after reaching a certain threshold) or displacement-driven (sudden jump in force when the threshold is reached). The intersection of SH and SS curves represents the *dual stiffness point*, at which the system exhibits two different stiffnesses corresponding to the same extension and tensile force, resulting in two different equilibrium configurations. This dual stiffness property can be harnessed to achieve unique dynamic behavior, which is analyzed in subsequent sections. The expression of non-dimensional potential energy can be expressed as follows (Eq. (3) and Fig. 2(b)):

$$\tilde{U} = \frac{\tilde{x}^2}{2} + \tilde{\kappa}\frac{\tilde{x}^4}{4} \quad (5)$$

Fig. 3(b) shows the potential energy variation corresponding to the SH and SS modes. Both potential curves have *stable equilibrium* points at the same displacements, which is also the *dual stiffness point*. The intersection of two potential wells represents the *mode switching* points where the system switches from one mode to another. The trajectories for the restoring force and the potential energy from the initial mode to the final mode can be shown in Fig. 3(c) and (d). The real-time rendering of the stiffness switching behavior in force-displacement and strain energy plots is provided in the supplementary material (Appendix B. Video S2). The system's dynamic response, exploring dual stiffness behavior and the CSI phenomenon, is provided in subsequent sections.

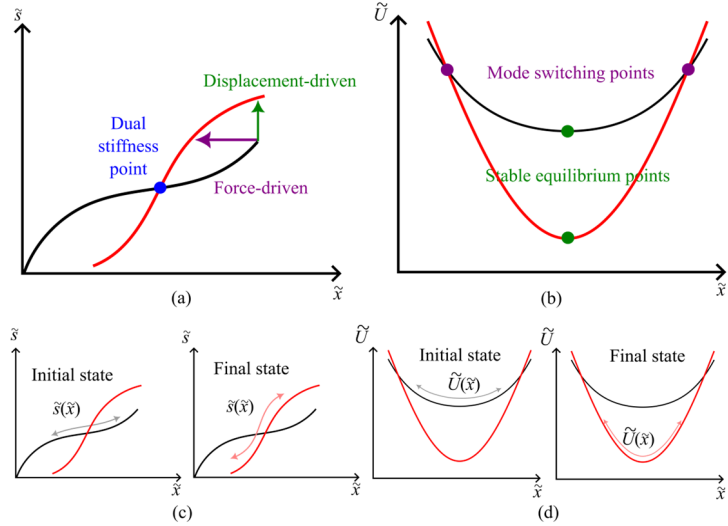

**Fig. 3.** (a) The counter-snap instability (CSI) behavior is represented from the force displacement profile. The stiffness switching can be obtained through *force-driven* or *displacement-driven* behavior. The intersection of strain hardening (SH) and strain softening (SS) curves represents the *dual stiffness point*, at which the system exhibits two different stiffnesses corresponding to the same extension and the same tensile force, resulting in two different equilibrium configurations. (b) The corresponding potential energy variation for the two stable modes. Both potential curves have *stable equilibrium points* at the same displacements, which is also the *dual stiffness point*. The intersection of two potential curves represents the *mode switching points* where the system switches from one mode to another. (c, d) Trajectories for the restoring force and the potential energy from the initial to the final mode.

## 5. Results and discussions

### 5.1. Dynamic response

The system's dynamic response obtained from the analytical solution of Eq. (4) is shown in Fig. 4. The values of the non-dimensional parameters used in the present study are given in Table 1. Fig. 4(a), represents a 3D plot illustrating the relationship between frequency ratio ($\lambda$), dimensionless time ($\tau$), and dimensionless displacement amplitude ($\tilde{x}$). The grey and pink dotted curves represent the system's strain hardening (SHFD) and strain softening (SSFD) frequency responses in the frequency-amplitude plane. The blue solid line captures the dynamic counter-snapping instability (CSI) due to the negative extensibility (NE) phenomenon, activated when the threshold amplitude, $\tilde{x}_T$ is exceeded. For details on the stable and unstable branches, refer the Fig. D.1 of the supplementary materials (Appendix D). Initially, the system vibrates in the SHFD; upon the threshold ($\tilde{x}_T = 1$), the stiffness switches to the SSFD regime, leading to an 86.59% reduction in displacement amplitude (from 1 to 0.134), thereby avoiding resonance. When the system again exceeds the threshold, it switches back to the SHFD regime, causing a further amplitude to drop from 1 to 0.17, corresponding to an 83% reduction.

Also, in Fig. 4(a), the time-amplitude plane shows four time-domain responses (I to IV) for $\lambda$ = 0.5, 1.1, 1.8 and 2.5, each illustrating the system behavior under varying excitations. The grey and pink solid lines ($\lambda$ = 0.5 and 2.5) represents stable vibrations in the strain hardening (SHTDR) and strain softening (SSTDR) regimes, with peak amplitude remaining below $\tilde{x}_T$. Near the natural frequencies ($\lambda$ = 1.1 and 1.8), the system exhibits switching behavior: black-to-red trajectories indicate transitions from SHTDR to SSTDR, while red-to-black trajectories denote the reverse, triggered as $\tilde{x}$ exceeds $\tilde{x}_T$. Fig. 4(b) displays the corresponding phase portraits for the same $\lambda$ values, showing mode-space trajectories. Grey and pink solid plots represent motions confined to the strain-hardening (SHPP) and strain-softening (SHPP) regions, respectively. Mixed-color trajectories highlight transitions, with black-to-red segments indicating SHPP to SSPP switching, and red-to-black segments indicating SSPP to SHPP. The following subsection explores the dependencies of the parameters given in Table 1 with the behavior of NE mechanism.

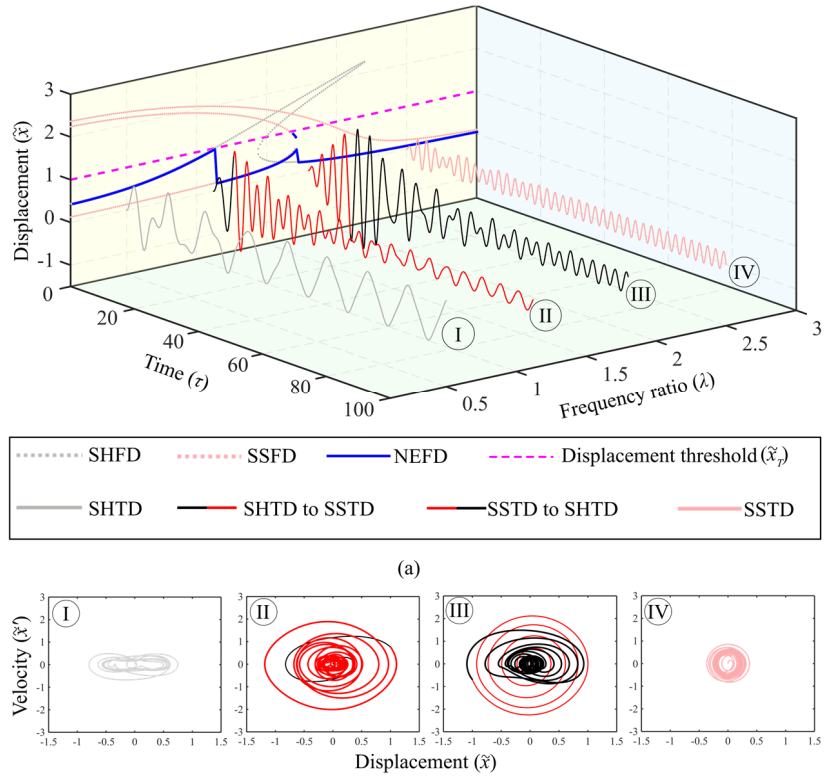

**Fig. 4.** (a) 3D plot showing the relationship between frequency ratio ($\lambda$), dimensionless time ($\tau$) and dimensionless displacement amplitude ($\tilde{x}$). In the frequency amplitude plane, the grey dotted curve represents strain hardening response (SHFD), the pink dotted curve denotes strain-softening response (SSFD), and the blue curve indicates dynamic negative extensibility (NE) when the threshold amplitude, $\tilde{x}_T$ (dashed magenta line) is exceeded. The time amplitude plane displays four time-domain responses (I to IV) for $\lambda$ = 0.5, 1.1, 1.8 and 2.5, where grey and pink solid lines correspond to strain hardening (SHTD) and strain-softening (SSTD) regimes, respectively. Mixed-color time trajectories highlight switching behavior: black-to-red indicates SHTD to SSTD, and red-to-black curve indicates SSTD to SHTD. (b) Corresponding phase portraits for the same $\lambda$ values illustrating the mode-space trajectories: grey for strain-hardening (SHPP), pink for strain-softening (SSPP), color transitions marking regime shifts (SHPP ↔ SSPP).

**Table 1** Non-dimensional parameters used in the present study.

| Parameters | Damping ratio | Stiffness ratio | Non-dimensional amplitude | Displacement threshold |
|---|---|---|---|---|
| Symbols | $\xi$ | $\tilde{\kappa}$ | $\Gamma$ | $\tilde{x}_T$ |
| Values | 0.05 | 1 or -0.25 | 0.5 | 1 |

### 5.2. Parametric space

From Fig. 4, it is observed that the occurrence of the NE phenomenon is primarily dependent on three parameters: (i) Amplitude of the excitation ($\Gamma$), (ii) Damping ratio ($\xi$), and (iii) Excitation frequency ($\lambda$). Thus, to obtain a more comprehensive visualization of the characteristics of the NE mechanism, a three-dimensional parametric space between $\Gamma$, $\xi$ and $\lambda$ is shown in Fig. 5 considering 100 × 100 × 100 grid. To obtain the parametric space, a time response of 100 forcing cycles is obtained for each simulation, considering the system first following the SH force displacement curve, as mentioned in the section 4. If the response amplitude, which begins at $\tilde{x} = 0$ based on the initial conditions, exceeds the displacement threshold $(\tilde{x}_T)$, then the CSI phenomenon occurs and the system switches from SH to SS. Alternatively, if the response amplitude $(\tilde{x})$ does not exceed the displacement threshold $(\tilde{x}_T)$, then we recognize that the CSI phenomenon did not occur, and the system remains in SH. Thus, among all possible combinations of $\Gamma$, $\lambda$, and $\xi$, the surface plots in Fig. 5 specifically enclose the domain of those combinations for which the CSI phenomenon has occurred.

In Fig. 5, three values of the non-dimensional linear stiffness for the SH $(\tilde{\beta})$ have been considered, i.e., $\tilde{\beta} = 0.5, 1$, and 2. The values of other non-dimensional parameters, such as the non-dimensional non-linear stiffness, for the SH $(\tilde{\alpha})$, non-dimensional linear stiffness for the SS $(\tilde{\delta})$ and non-dimensional non-linear stiffness for the SS $(\tilde{\gamma})$ are chosen as $\tilde{\alpha} = \tilde{\beta}$, $\tilde{\delta} = 4\tilde{\beta}$ and $\tilde{\gamma} = -\tilde{\beta}$. The reason behind choosing these values is to replicate the spring paradox system given in the section 2. Also, by choosing these values, we can maintain the uniformity of the term stiffness ratio $(\tilde{\kappa})$ as $\tilde{\kappa} = 1$ for SH and $-0.25$ for SS. It is observed that, when $\tilde{\beta} = 0.5$, the surface of the parametric space considers a larger combination of $\Gamma$, $\lambda$, and $\xi$. In other words, the CSI phenomenon occurs for even low values of $\Gamma$ and higher values of $\xi$. However, when the value of $\tilde{\beta}$ is increased up to 1, the parametric space reduces and shifts towards higher values of $\lambda$. The reduction of the parametric space represents the regions where the CSI phenomenon occurs, but only for high amplitudes and low damping. This is due to the overall increase in the stiffness of the overall system. Hence, the high-amplitude/low-damping combinations highlighted above represent the minimum excitation and dissipation levels necessary to trigger the negative-extensibility mechanism; without meeting these thresholds, the CSI response does not materialize. Now, when the value of $\tilde{\beta}$ is increased up to 2, the surface further reduces and also shifts to a higher $\lambda$. This is due to the change in the system's natural frequency as $\tilde{\beta}$ is directly proportional to the initial stiffness of the system. Due to the increase in stiffness, combinations of very high amplitude $(\Gamma)$ and very low damping $(\xi)$ are required to induce the NE mechanism. A detailed variation of the surface plot with the values of $\tilde{\beta}$ is shown in Fig. E.2 of the supplementary materials (Appendix E). The following subsection explores the effectiveness of NE systems compared to bi-stable systems.

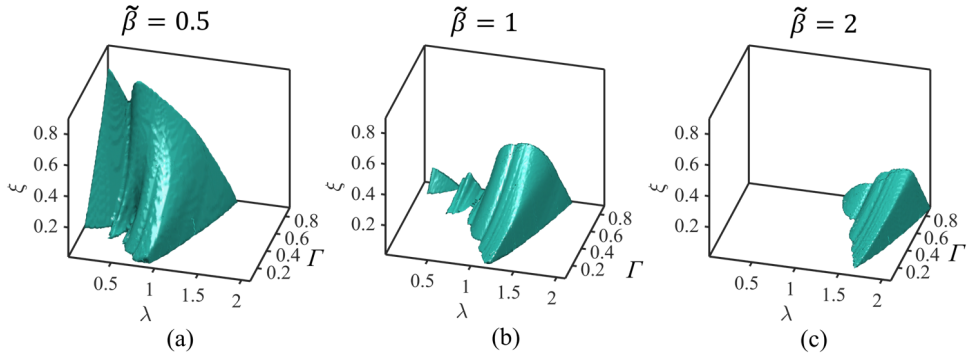

**Fig. 5**. Three-dimensional forcing amplitude-excitation frequency-damping ratio ($\Gamma$ vs $\lambda$ vs $\xi$) parametric space where the surface plot represents the volume of all combinations of $\Gamma$, $\lambda$ and $\xi$ for which the negative extensibility phenomenon is achieved for (a) $\tilde{\beta} = 0.5$, (b) $\tilde{\beta} = 1$ and (c) $\tilde{\beta} = 2$.

*5.3. Comparison with Bi-stable systems*

The efficacy of the dynamic performance of the proposed NE model is compared with existing bi-stable systems. When a structural system exhibits two statically stable configurations, it is said to be bistable [25]. In a bi-stable mechanism, the system switches from one stable region to another stable region when a certain amount of displacement threshold is exceeded. This bi-stable behavior can be obtained in several ways, such as using the Duffing Oscillator [26,27], beam with magnetic arrangements [28,29], Kresling Origami [30,31], etc. Typically, the behavior of a bi-stable system is categorized as symmetric or asymmetric depending on the configuration. The basic difference between symmetric and asymmetric bi-stable systems lies in the behavior of the force displacement plot, i.e., in asymmetric bi-stable systems, the linear stiffness of the first stable mode is different from the second stable mode [32,33]. It is observed that asymmetric bi-stable system outperforms symmetric bi-stable systems by achieving smaller and larger natural frequencies at two positions, offering the capability to adjust the equilibrium position in response to excitation of varying frequencies, thereby further mitigating the impact of vibration [28,30].

Thus, the dynamic performance of the proposed model is compared with two types of asymmetric bi-stable systems. The force-displacement and potential energy curves for two different types of asymmetric bi-stable systems are shown in Fig. 6. The key difference between the two systems lies in the shape of the potential wells. In the *Type I* case (see Fig. 6(b)), the left well is *shifted* so that its minimum is at a different potential value than the other making the system a built-in bias (one minimum is energetically favored). Whereas, for the *Type II* (see Fig. 6(d)), the two minima of the wells sit at the same potential energy (i.e., both at zero), making the system unbiased or *degenerate* in terms of energy. However, the left width of the potential wells for the *Type I* case (see Fig. 6(d)) is wider than the right width, representing lower linear stiffness than the right well. The present negative extensibility (NE) model compares its dynamic performance with the bi-stable systems mentioned. The force-displacement and potential energy curves are non-dimensionalized for effective comparison and fitted by a *quintic* polynomial to facilitate subsequent dynamic analysis. The expression for non-dimensional force-displacements is given as follows:

$$\tilde{\kappa}_i = \sum_{q=1}^{5} \eta_{iq} \tilde{x}_i^q \qquad (6)$$

where, $\tilde{\kappa}_i$, $\tilde{x}_i$ and $\eta_{iq}$ represents the non-dimensional restoring force, displacements, and polynomial coefficients for either the *Type I* system ($i = I$) or *Type II* system ($i = II$). The values of $\eta_{iq}$ for the case of *Types I* and *II* are given in Table 2. Now, for effective comparison with the present NE system, the non-dimensional force displacement profile defined in Fig. 3(a) is modified in such a manner that the non-dimensional linear stiffness of strain hardening (SH) $(\tilde{\beta})$ is kept equal with the non-dimensional linear stiffness of the first stable state of the bi-stable system $(\tilde{\beta}_I$ or $\tilde{\beta}_{II})$ as shown in Fig. 7. Similarly, the non-dimensional linear stiffness of strain softening (SS) $(\tilde{\delta})$ is kept equal with the non-dimensional linear stiffness of the second stable state of the bi-stable system $(\tilde{\delta}_I$ or $\tilde{\delta}_{II})$. The non-dimensional threshold displacement for both the cases of negative extensibility (NE) system $(\tilde{x}_I^T$ or $\tilde{x}_{II}^T)$ is chosen based on the distance between the equilibrium point of the first stable state to the unstable equilibrium point of the bi-stable system. The four linear stiffness values and the two displacement threshold values used in the present study are given in Table 3.

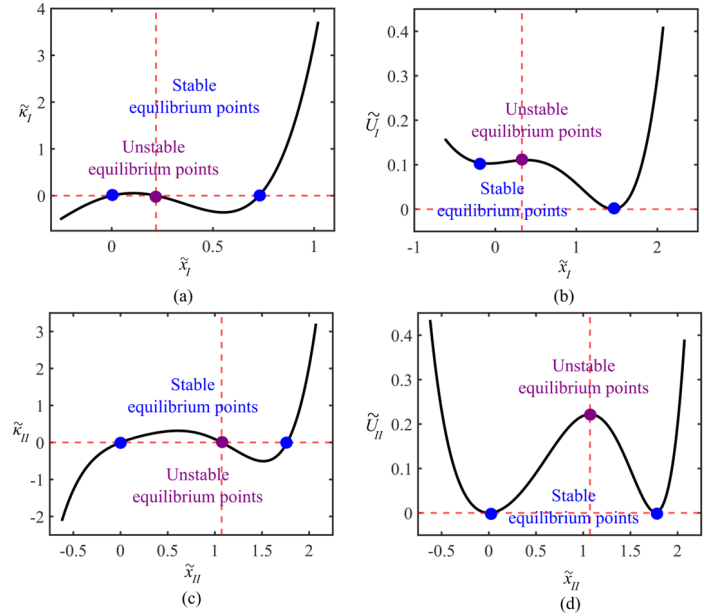

**Fig. 6.** Force-displacement profiles and potential energy curves for (a) and (b) Bi-stable system having left well shifted so that its minimum is at a different potential value than the other (*Type I*), (c) and (d) Bi-stable system having the two minima of the wells sit at the same potential energy (*Type II*). However, the left width of the potential wells for the degenerate case is wider than the right width, representing lower linear stiffness than the right well.

**Table 2** The fitting coefficients of the non-dimensional force displacement expression given in Eq. (6)

| Coefficients | $\eta_{i1}$ | $\eta_{i2}$ | $\eta_{i3}$ | $\eta_{i4}$ | $\eta_{i5}$ |
|---|---|---|---|---|---|
| $\eta_{Iq}$ | 1 | -4 | -1 | 7 | 1 |
| $\eta_{IIq}$ | 1 | -1 | 2 | -3 | 1 |

**Table 3** Non-dimensional parameters used for comparing asymmetric bi-stable systems (*Type I* and *II*) with the NE system.

| Comparison Parameters | $\tilde{\beta}_I$ | $\tilde{\beta}_{II}$ | $\tilde{\delta}_I$ | $\tilde{\delta}_{II}$ | $\tilde{x}_I^T$ | $\tilde{x}_{II}^T$ |
|---|---|---|---|---|---|---|
| Values | 1 | 1 | 2.6 | 15 | 0.25 | 1 |

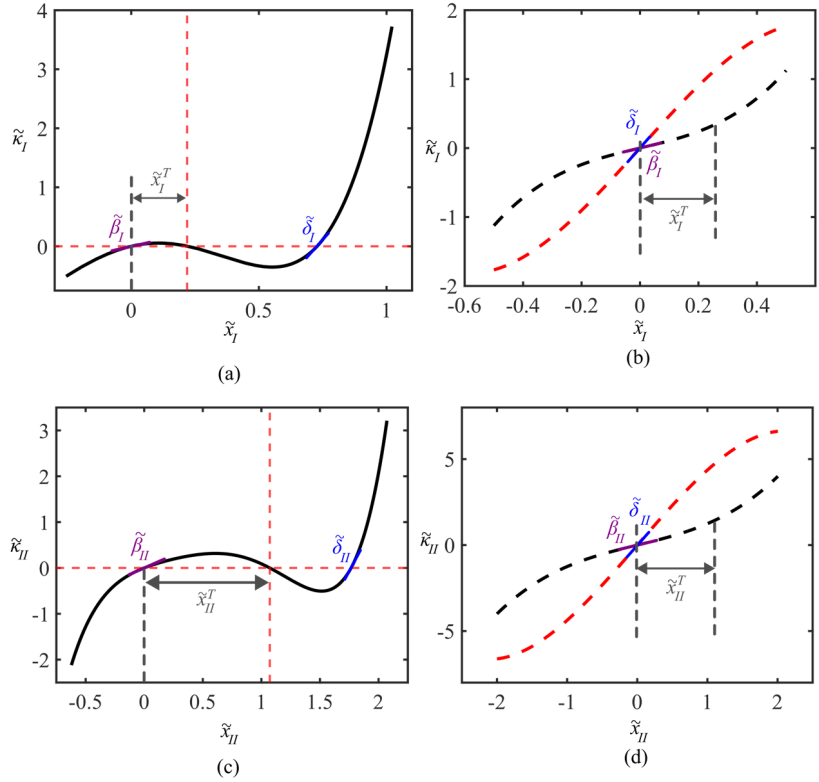

**Fig. 7**. Force-displacement curves of the negative extensibility system in compared with (a) and (b) *Type I* and (c) and (d) *Type II* systems. For effective comparison, the non-dimensional linear stiffness of strain hardening (SH) $(\tilde{\beta})$ is kept equal with the non-dimensional linear stiffness of the first stable state of the bi-stable system ($\tilde{\beta}_I$ or $\tilde{\beta}_{II}$). Similarly, the non-dimensional linear stiffness of strain softening (SS) $(\tilde{\delta})$ is kept equal with the non-dimensional linear stiffness of the second stable state of the bi-stable system ($\tilde{\delta}_I$ or $\tilde{\delta}_{II}$). The non-dimensional threshold displacement for both the cases of negative extensibility (NE) system ($\tilde{x}_I^T$ or $\tilde{x}_{II}^T$) is chosen based on the distance between the equilibrium point of the first stable state to the unstable equilibrium point.

Now, the dynamic performance of the proposed model is compared with the two bi-stable systems using time-domain response curves shown in Fig. 8. Fig. 8(a)-(c) compares the proposed NE with the *Type I* system. Similarly, Fig. 8(d)-(f) compares the proposed NE with the *Type II* system. From all the twelve-time domain responses, three figures of the first and third row represent the results of asymmetric bi-stable systems (*Type I and II*), and the second and fourth rows represent the time domain responses of the proposed NE systems. Three different values of excitation amplitudes are chosen for each bi-stable system to show the different behavior, such as *Intra-well* for $\Gamma$ = 0.01 and 0.13, *Intermittent Cross-Well* or *Aperiodic Vacillating* behavior for $\Gamma$ = 0.075 and 0.485 and *Continuous Cross-Well* or *Periodic Vacillating* behavior for $\Gamma$ = 0.1 and 0.5. More details on these types of behavior can be found in [26,34,35].

In Fig. 8, the black color → dynamic response in the first stable state (left well in bi-stable / strain hardening in NE); whereas, the red color → dynamic response in the second stable state (right well in bi-stable / strain softening in NE). From Fig. 8(a) and (d), it is observed that when the excitation amplitude is significantly less (*Intra-Well* with $\Gamma$ = 0.01 for *Type I* and $\Gamma$ = 0.13 for *Type II*), the maximum steady state dynamic response of both asymmetric bi-stable systems is similar to the proposed NE system, i.e., approximately $\tilde{x}_{I_{max}} \approx \tilde{x}_{max} \approx 0.2$ when comparing with *Type I* systems and approximately $\tilde{x}_{II_{max}} \approx \tilde{x}_{max} \approx 1.2$ when comparing with *Type II* systems. This is mainly due to the system's vibration in the linear region due to a lower amplitude. Since the linear stiffness for the bi-stable and proposed NE systems is the same, the maximum displacement amplitude is also the same.

However, when the excitation amplitude increases slightly, i.e., when *Intermittent Cross-Well* or *Aperiodic Vacillating* behavior is observed ($\Gamma$ = 0.075 for *Type I* and $\Gamma$ = 0.485 for *Type II*) as shown in Fig. 8(b) and (e), the maximum dynamic response for both the bi-stable systems is significantly high in the transient phase ($\tilde{x}_{I_{max}} \approx 1.05$ and $\tilde{x}_{II_{max}} \approx 3.68$) compared to the steady state response ($\tilde{x}_{I_{max}} \approx 0.14$ and $\tilde{x}_{II_{max}} \approx 1.02$). Meanwhile, the proposed NE system has a significant reduction in the dynamic response, both in transient ($\tilde{x}_{max} \approx 0.61$) and steady state phase ($\tilde{x}_{max} \approx 0.1$) when compared to *Type I* (Fig. 8(b)) and $\tilde{x}_{max} \approx 2$ in the transient phase and $\tilde{x}_{max} \approx 0.37$ in steady state, when compared with *Type II* (Fig. 8(e)). This leads to a 42% reduction in transient response and 28.57% in steady state response compared to *Type I*. Similarly, a reduction of 45% in transient response and 63.72% in steady state response is observed when compared with *Type II*.

Now, if the excitation amplitude is further increased, i.e., $\Gamma$ = 0.1 for *Type I* and $\Gamma$ = 0.5 for *Type II*, which shows the *Continuous Cross-Well* or *Periodic Vacillating* behavior in bi-stable systems, as shown in Fig. 8(c) and (f), the maximum dynamic response for both the bi-stable systems (*Type I and II*) is significantly high in both transient and steady state responses ($\tilde{x}_{I_{max}} \approx 1.28$ and $\tilde{x}_{II_{max}} \approx 3.55$). However, for the same excitation amplitude, the proposed NE behavior shows a substantial reduction in maximum dynamic response, i.e., $\tilde{x}_{max} \approx 0.13$ (89.84%) reduction when compared with *Type I* and $\tilde{x}_{max} \approx 0.37$ (89.57%) reduction when compared to *Type II*. Thus, if the excitation amplitude is high in the bi-stable system, the system goes into *inter-well oscillations*, resulting in increased dynamic response. However, the proposed NE system will have reduced dynamic response even if the excitation amplitude is high enough. For the time domain response shown in Fig. 8, corresponding phase portraits are also compared between the proposed system and bi-stable systems (*Type I* and *II*). Detailed results are provided in the supplementary materials, Fig. F.3 (Appendix F). Similar to the time domain response, the phase portraits also show a significant reduction in the dynamic response, both in the transient and steady state phases.

For comprehensive visualization, a real-time rendering of the time domain response and the phase portraits comparing the proposed system with bi-stable systems (*Type I* and *II*) is provided in the supplementary material (Appendix G. Video S3). Further investigations were conducted to evaluate the efficacy of the proposed NE system in comparison with asymmetric bi-stable systems (*Type I* and *II*) under high-amplitude vibrations, particularly for cases exhibiting *Continuous Cross-Well* or *Periodic Vacillating* behavior. The study considered non-dimensional forcing amplitudes of $\Gamma = 0.125, 0.15$, and $0.175$ for the *Type I* bi-stable and $\Gamma = 1, 2$, and $3$ for the *Type II* bi-stable system. For detailed results, refer to Fig. H.4 in the supplementary materials (Appendix H). The findings reveal that the proposed NE system exhibits approximately 80–90% better performance in transient and steady-state displacement amplitudes under high-amplitude excitations, irrespective of the bi-stable system type.

Moreover, additional comparisons have also been conducted for the case of switching from strain softening (SS) to strain hardening (SH) in the proposed NE system, in contrast to switching from high- to low-stiffness potential wells in bi-stable systems (*Type I* and *II*). Detailed results are provided in the supplementary materials, Fig. I.5 (Appendix I). These findings also indicate that the proposed NE system offers approximately 73–77% improved vibration isolation compared to either type of bi-stable system in the SS-to-SH switching scenario.

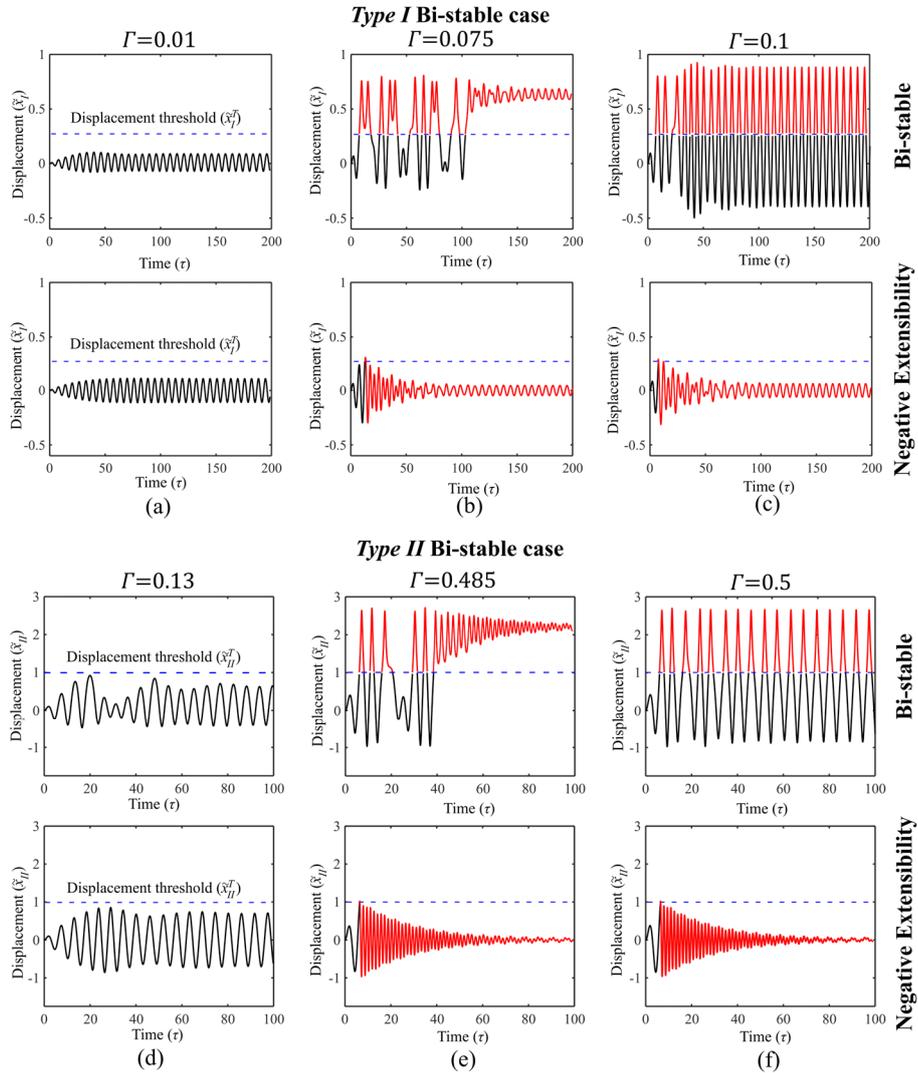

**Fig. 8**. Time domain response curves comparing the dynamic performance of the proposed model with two types of bi-stable systems having asymmetric potential wells for various excitation amplitudes ($\Gamma = 0.01, 0.075, 0.1, 0.13, 0.485, 0.5$). (a)-(c) Comparison with *Type I* bi-stable. (d)-(f) Comparison with *Type II* bi-stable. The black solid lines represent the dynamic response for the system existing in the first stable state (left potential well of the bi-stable system and strain hardening case for the NE system). Similarly, the red solid lines represent the dynamic response for the system existing in the second stable state (right potential well of the bi-stable system and strain softening case for the NE system).

# 6. Conclusions

The present study investigates the dynamics of a generalized negative extensibility system inspired by the spring (Braess) paradox concept. This study focuses on the dynamics of a generalized single spring mass damper system with tunable nonlinear spring stiffness under harmonic force excitation. We observed that the system's stiffness behavior switches from strain-hardening to strain-softening once a certain displacement threshold is exceeded. This passive stiffness switching enables resonance avoidance through countersnapping instability.

Additionally, the activation of the negative extensibility mechanism is highly sensitive to the isolator's damping, as well as to the amplitude and frequency of the excitation. This sensitivity is analyzed within a three-dimensional parametric space, which predominantly spans regions of low damping, high amplitude, and excitation frequencies near the system's natural frequency. As the non-dimensional stiffness increases, the surface area of activation shrinks and shifts toward higher non-dimensional excitation frequencies, owing to its direct proportionality with the system's initial stiffness.

Compared to bi-stable systems with identical baseline stiffness characteristics, the proposed system achieves superior vibration isolation due to controllable mode switching and stiffness modulation without drastically changing the overall geometry of the system. The considered system demonstrates around 80 to 90% better performance in steady-state displacement amplitude under higher amplitude vibrations, regardless of the bistable system type. Although theoretical results are promising, practical realization will still require future development.

Future studies could investigate the system's performance under random vibrations or harsh environments such as earthquakes, impact loads, etc. Additionally, examining the effectiveness of the proposed system in multiple directions simultaneously would be beneficial. The exploration of countersnapping dynamics for vibration control is in its early stages, yet we anticipate that, in time, countersnapping instabilities will rival conventional snapping instabilities in their practical impact.

**Acknowledgments**

We acknowledge the support by the National Research Foundation grants funded by the Korea government [Grants No. 2023R1A2C2003705 and No. 2022H1D3A2A03096579 (Brain Pool Plus by the Ministry of Science and ICT)] and the Korea Research Institute for defense Technology planning and advancement (KRIT) grant funded by the Defense Acquisition Program Administration (DAPA) [Grant No. 20-105-E00-005 (KRIT-CT-23-010), VTOL Technology Research Center for Defense Applications, 2025].